\documentclass{ws-procs961x669}            
\usepackage{amsmath}
\usepackage{amssymb}
\usepackage{hyperref}
\usepackage{subfigure}
\usepackage{bm}
\usepackage{aas_macros}
\usepackage{orcidlink}
\usepackage{physics}

\hypersetup{
	colorlinks=true,
	linkcolor=blue,
	filecolor=magenta,
	urlcolor=red,
	citecolor=magenta,
	pdftitle={Sharelatex Example},
}

\begin{document}
\title{Asymptotically flat black hole solution in modified gravity}

\author{Surajit Kalita$^1$\orcidlink{0000-0002-3818-6037} and Banibrata Mukhopadhyay$^2$\orcidlink{0000-0002-3020-9513}}

\address{Department of Physics, Indian Institute of Science, Bangalore-560012, India\\
$^1$E-mail: surajitk@iisc.ac.in\\
$^2$E-mail: bm@iisc.ac.in}

\begin{abstract}
From time to time, different observations suggest that Einstein's theory of general relativity (GR) may not be the ultimate theory of gravity. Various researchers have suggested that the $f(R)$ theory of gravity is the best alternative to replace GR. Using $f(R)$ gravity, one can elucidate the various unexplained physics of compact objects, such as black holes, neutron stars, and white dwarfs. Researchers have already put effort into finding the vacuum solution around a black hole in $f(R)$ gravity. However, for a long time, they could not find an asymptotically flat vacuum solution. In this article, we show that the asymptotically flat vacuum solution of $f(R)$ gravity is possible and thereby use it to explain the spherical accretion flow around the black hole.
\end{abstract}

\keywords{black hole, modified gravity, spherical accretion.}

\bodymatter

\section{Introduction}
Einstein's theory of general relativity (GR) is undoubtedly the most beautiful theory to explain gravity. It can easily explain a large number of phenomena where Newtonian gravity falls short, such as the deflection of light in strong gravity, generation of the gravitational waves, perihelion precession of Mercury's orbit, gravitational redshift of light, to mention a few. However, following several recent observations in cosmology and the high-density regions in the universe, such as the vicinity of compact objects, it seems that GR may not be the ultimate theory of gravity. Starobinsky first used one of the modified theories of gravity, namely $R^2$-gravity \cite{1980PhLB...91...99S}, to explain the cosmology of the very early universe. Eventually, researchers have proposed different modifications of GR, e.g., various models of $f(R)$ gravity, to elaborate on various unexplained physics related to compact objects. Capozziello and his collaborators used different forms of $f(R)$ gravity to explain the massive neutron stars \cite{2014PhRvD..89j3509A}. Similarly, Kalita \& Mukhopadhyay used $f(R)$ gravity to unify the physics of all white dwarfs, including those possessing sub- and super-Chandrasekhar limiting masses \cite{2018JCAP...09..007K}, and show that they can be detected by various proposed gravitational wave detectors \cite{2021ApJ...909...65K}.

According to Birkhoff's theorem, the spherically symmetric vacuum solution in GR is static, which is the famous Schwarzschild solution. In the same context, the vacuum solution of $f(R)$ gravity is also an interesting problem. Multam\"{a}ki \& Vilja first obtained the solutions for a static, spherically symmetric vacuum spacetime in $f(R)$ gravity \cite{2006PhRvD..74f4022M}, and argued that for a large class model, Schwarzschild-de Sitter metric is an exact solution of the field equations. Eventually, many researchers have obtained several solutions for different modified theories of gravity in various spacetime geometries. Capozziello and his collaborators obtained spherically symmetric as well as axially symmetric vacuum solutions in $f(R)$ gravity using Noether symmetry \cite{2007CQGra..24.2153C,2010CQGra..27p5008C}. These new solutions alter the black hole's event horizon, thereby various critical orbits, such as marginally stable, marginally bound, and photon orbits. In this way, modified theories of gravity change the dynamics of the particles moving around the black hole.

The above-mentioned solutions, introduced in these literatures, always contain some diverging terms in the metric components. Hence these solutions never reduce to the Minkowski metric at the asymptotically flat limit. This asymptotic flatness is essential in the context of physical problems such as accretion disc, as a disc is expected to extend in a vast region around the compact object. Therefore, at the outermost portion of the disc, which is very far from the compact object, no physics should be violated as given by the Schwarzschild/Minkowski metric. Many of the models mentioned above assume scalar curvature, $R \neq 0$ throughout the spacetime, which is again questionable as for the Schwarzschild metric, $R=0$, which needs to be satisfied at the asymptotic flat limit. Kalita \& Mukhopadhyay, for the first time, obtained the spherically symmetric asymptotically flat vacuum solution in the premise of $f(R)$ gravity \cite{2019EPJC...79..877K}. In this article, we primarily discuss this solution and its applications in the spherical accretion flow around a black hole.

\section{Basic formalism}
The 4-dimensional action for $f(R)$ gravity in vacuum is given by \cite{2010LRR....13....3D}
\begin{align}
	S = \frac{c^4}{16 \pi G}\int \sqrt{-g}~f(R)\dd[4]{x},
\end{align}
where $c$ is the speed of light, $G$ is Newton's gravitational constant, $R$ is the scalar curvature, and $g=\det(g_{\mu\nu})$ is the determinant of the metric $g_{\mu\nu}$. Varying this action with respect to $g_{\mu\nu}$, with appropriate boundary conditions, we obtain the modified Einstein equation for the $f(R)$ gravity, given by
\begin{equation}\label{modified Einstein}
	f'(R)G_{\mu \nu}+\frac{1}{2}g_{\mu \nu}\left[Rf'(R)-f(R)\right]-\left(\nabla_\mu \nabla_\nu-g_{\mu \nu}\Box\right)f'(R) = 0,
\end{equation}
where $\Box$ is the d'Alembertian operator given by $\Box=\nabla^\mu\nabla_\mu$ and $\nabla_\mu$ is the covariant derivative. For $f(R)=R$, Equation~\eqref{modified Einstein} reduces to the well-known Einstein field equation in GR.

\section{Asymptotically flat vacuum solution in $\bm{f(R)}$ gravity}
To obtain the spherically symmetric solution of the modified Einstein equation, we first assume $g_{\mu\nu} = \text{diag}(s(r), -p(r), -r^2, -r^2\sin^2\theta)$, where $s$, $p$ are the functions of the radial coordinate $r$ only. We know that far from the compact objects, the metric should reduce first to the Schwarzschild metric, and at very far from the compact object, it should reduce to the Minkowski metric as the effect of gravity is negligible there. Using these conditions at the asymptotic limit, one can show that for $f'(R) = 1+\beta/r$ with $\beta$ being a constant, in $c=G=M=1$ unit, the temporal and radial components of the metric are given by \cite{2019EPJC...79..877K}
\begin{align}\label{gtt}
	g_{tt} = s(r) = 1-\frac{2}{r}-\frac{\beta(\beta-6)}{2 r^2}+\frac{\beta^2 (13\beta-66)}{20 r^3}-\frac{\beta^3 (31\beta-156)}{48 r^4}+\dots,
\end{align}
\begin{align}\label{grr}
	g_{rr} = -p(r) = -\frac{16r^4}{(\beta+2r)^4}\frac{1}{s(r)}.
\end{align}
It is, of course, obvious from the above expressions that as $\beta \to 0$, it reduces to the Schwarzschild metric, and as $r\to \infty$, it reduces to the Minkowski metric, i.e., the flat spacetime.

\begin{figure}[!htbp]
	\centering
	\subfigure[temporal component]{\includegraphics[scale=0.38]{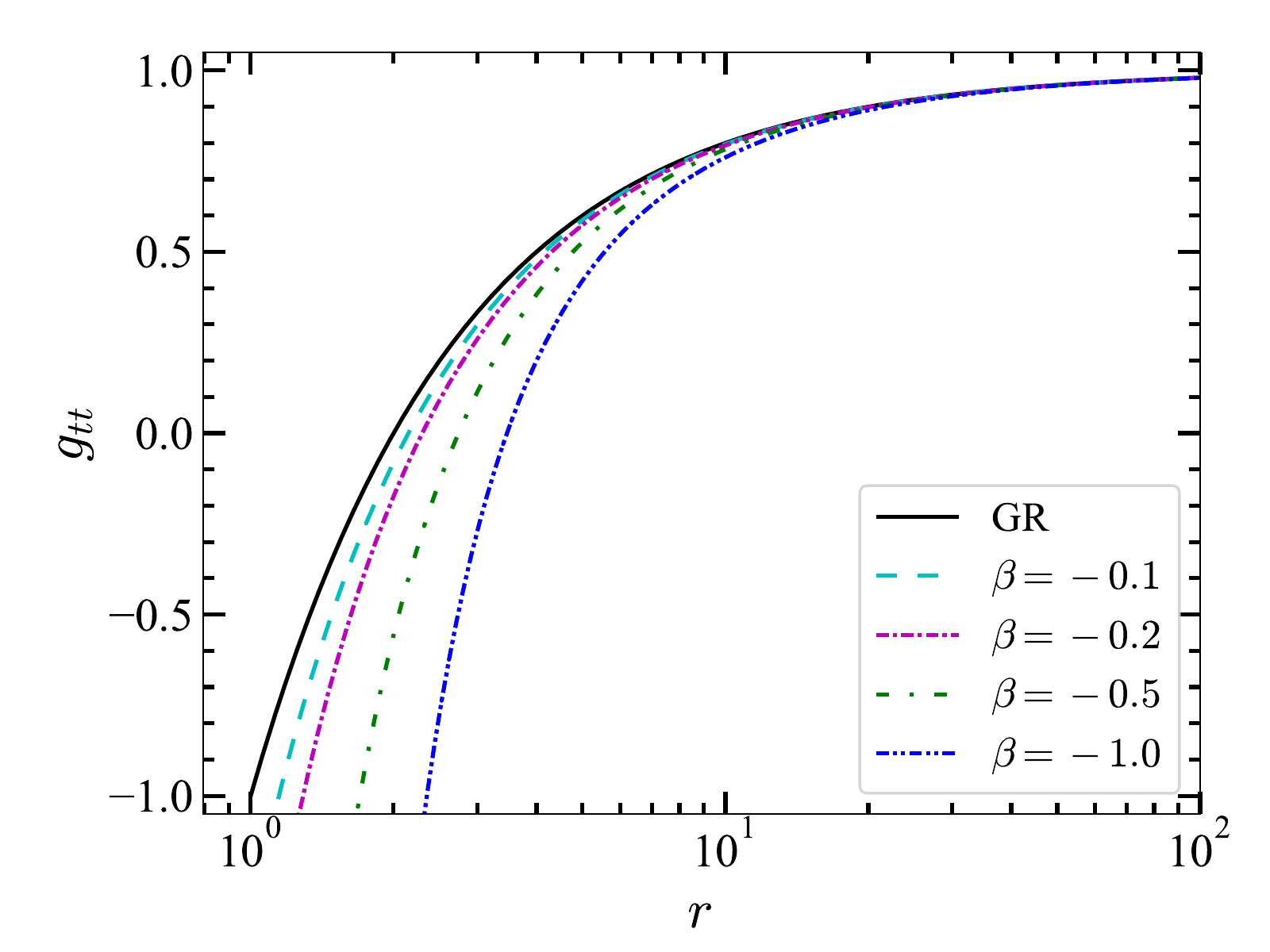}}
	\subfigure[radial component]{\includegraphics[scale=0.38]{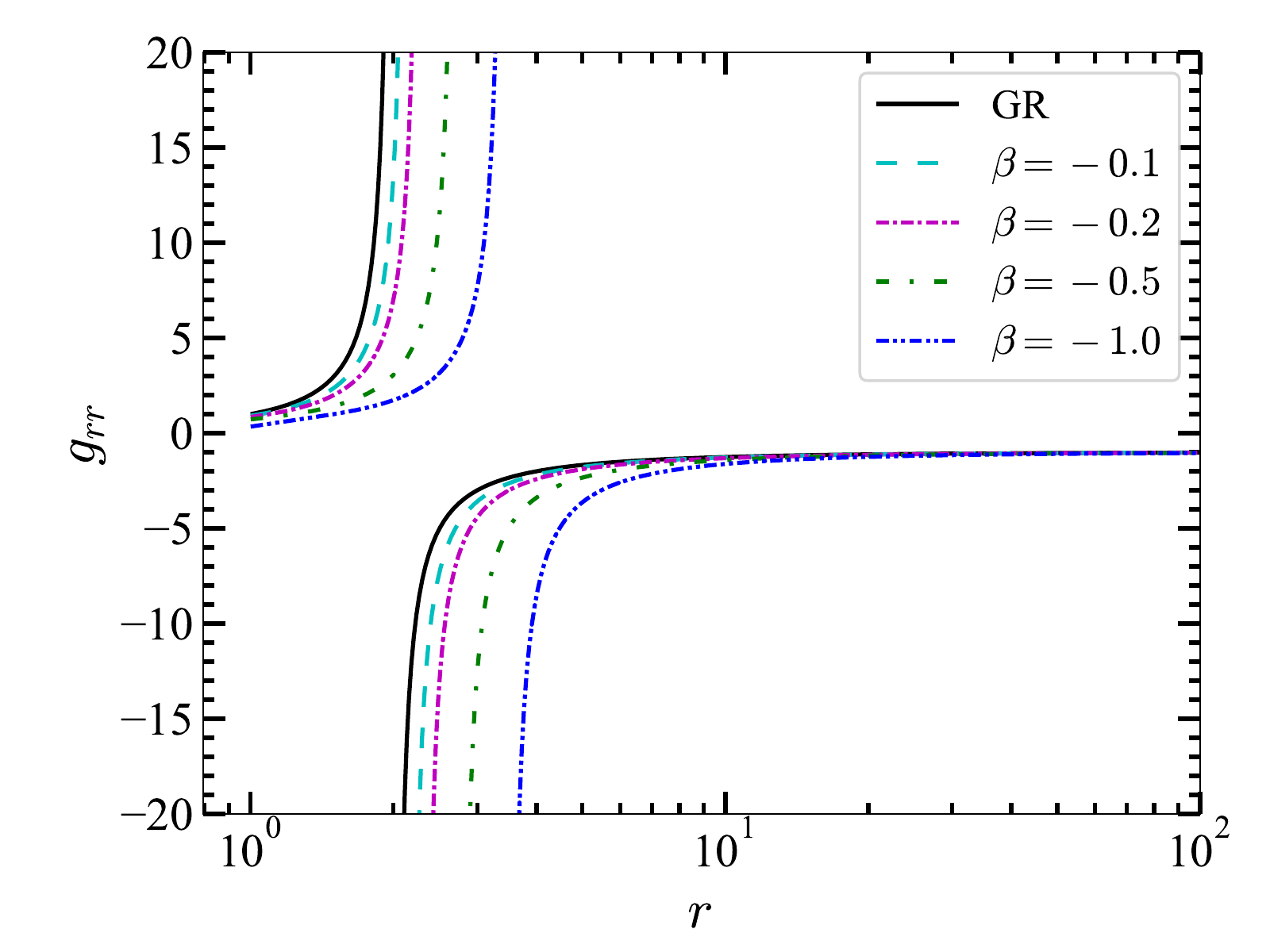}}
	\caption{Variation of temporal and radial metric elements as functions of distance $r$. All the quantities are in dimensionless units of $c=G=M=1$.}
	\label{temporal and radial}
\end{figure}

From Equations~\eqref{gtt} and~\eqref{grr}, it is observed that in $f(R)$ gravity, the vacuum solution is different from the Schwarzschild metric. This deviation indicates the violation of Birkhoff's theorem, which states that any spherically symmetric vacuum solution is essentially the Schwarzschild solution. Figure~\ref{temporal and radial} describes the variations of the metric elements $g_{tt}$ and $g_{rr}$ with respect to $r$ for different $\beta$. It is evident from the figure that as $r\to \infty$, the solutions reduce to the flat Minkowski metric, which, however, deviates near the black hole. $\beta = 0$ represents the Schwarzschild solution, with the radius of the event horizon $r_\mathrm{H} = 2$. In $f(R)$ gravity, as $\beta$ deviates from 0, the solution is no longer the Schwarzschild solution, and the physics related to these solutions alter. For example, one can easily obtain the event horizon from Equations~\eqref{gtt} and~\eqref{grr} by solving $g_{rr}\to \infty$, which shows that $r_\mathrm{H}$ shifts from 2 as $\beta$ deviates from 0. Not only the radius of the event horizon, but one can also obtain the various marginal orbits in $f(R)$ gravity from Equations~\eqref{gtt} and~\eqref{grr}. Table~\ref{parameters of space-time} shows the radii of various orbits and their critical angular momenta for different $\beta$. We choose $\beta$ to be negative because it guarantees that gravity is always attractive.

\begin{table}
	\tbl{Various characteristic parameters of spacetime for different values of $\beta$: $r_\mathrm{H}$ is the event horizon, $r_\mathrm{MB}$ the marginally bound orbit, $r_\mathrm{MS}$ the marginally stable orbit, $L_\mathrm{MB}$ and $L_\mathrm{MS}$ are their corresponding specific angular momenta, and $r_\mathrm{ph}$ is the photon orbit. All the values are in dimensionless unit considering $c=G=M=1$.}
	{\begin{tabular}{@{}lllllll@{}}
			\toprule
			$\beta$ & $r_\mathrm{H}$ & $r_\mathrm{MB}$ & $r_\mathrm{MS}$ & $r_\mathrm{ph}$ & $L_\mathrm{MB}$ & $L_\mathrm{MS}$\\
			\colrule
			0 (GR) & 2.00 & 4.00 & 6.00 & 3.00 & 4.00 & 3.46\\ 
			-0.1 & 2.15 & 4.30 & 6.45 & 3.20 & 4.15 & 3.61\\ 
			-0.2 & 2.30 & 4.60 & 6.90 & 3.40 & 4.29 & 3.75\\ 
			-0.5 & 2.74 & 5.52 & 8.28 & 3.98 & 4.71 & 4.15\\ 
			-1.0 & 3.47 & 7.07 & 10.64 & 4.94 & 5.37 & 4.78\\ 
			-1.5 & 4.18 & 8.66 & 13.08 & 5.89 & 5.99 & 5.37\\ 
			\botrule
		\end{tabular}
	}
	\label{parameters of space-time}
\end{table}

\section{Spherical accretion flow in $\bm {f(R)}$ gravity}
\begin{figure}[!htbp]
	\centering
	\subfigure[$T_\mathrm{out}=10^4\,$K]{\includegraphics[scale=0.38]{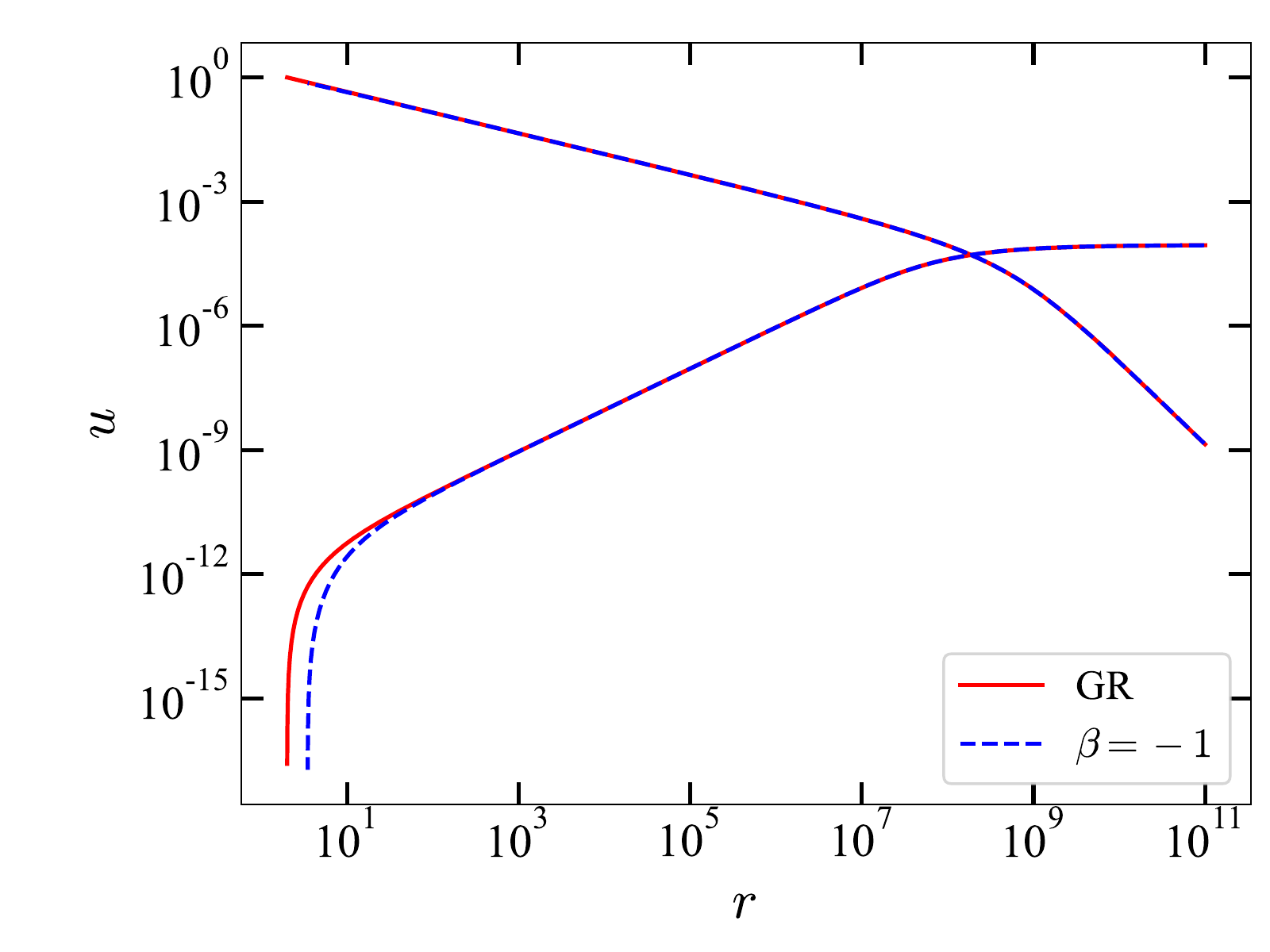}}
	\subfigure[$T_\mathrm{out}=10^{11}\,$K]{\includegraphics[scale=0.38]{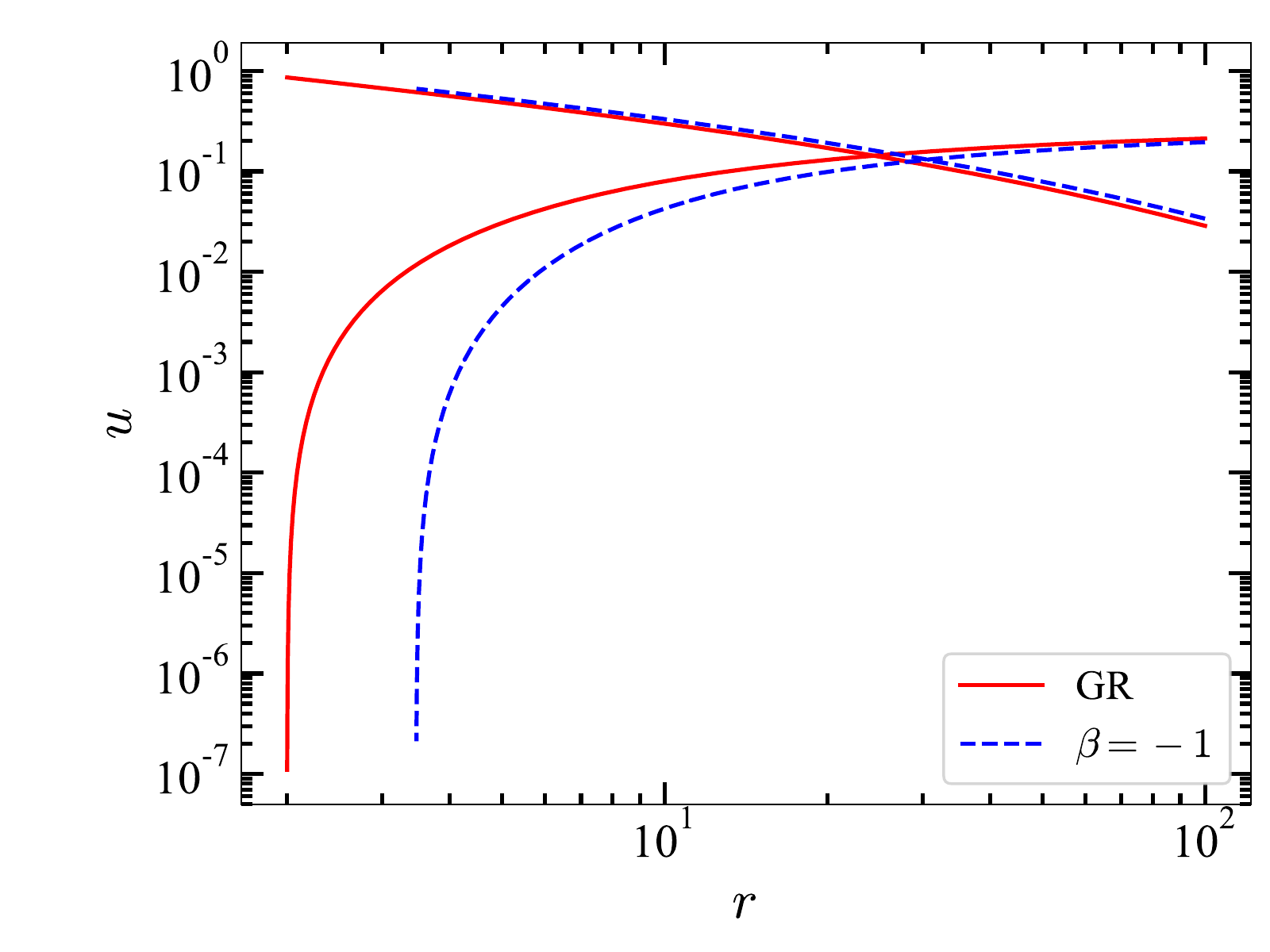}}
	\caption{Spherical accretion flow in modified gravity: red solid line corresponds to the Schwarzschild spacetime and blue dashed line corresponds to the $f(R)$ gravity with $\beta =-1$.}
	\label{bondi_plot}
\end{figure}

The vacuum solution for the $f(R)$ gravity can be used to understand the physics of black hole accretion. Here we explore the simplest class of it, i.e., spherical accretion flow around a black hole. In the Newtonian framework, such a flow is known as the Bondi accretion flow \cite{1952MNRAS.112..195B}. Since this flow is spherically symmetric, it does not possess any angular momentum, i.e., the flow is not rotating. The equations for conservation of energy and mass flux are respectively given by
\begin{align}
	T^{\mu}_{0;\mu} = 0, \quad {\rm and} \quad \left(\rho U^{\mu}\right)_{;\mu} = 0,
\end{align}
where $T^\mu_\nu$ is the stress-energy tensor, $\rho$ is the density, and $U^\mu$ is the four-velocity of the fluid. Using these equations, for a spherically symmetric metric, $g_{\mu\nu} = \text{diag}(-e^{2\phi(r)}, e^{2\lambda(r)}, r^2, r^2\sin^2\theta)$, the radial velocity ($u$) gradient equation for such flow is given by \cite{2019EPJC...79..877K,1972Ap&SS..15..153M}
\begin{equation}\label{Eq: Bondi}
	\frac{\dd{u}}{u}\left[V^2-\frac{u^2}{u^2+e^{-2\lambda}}\right]+\frac{\dd{r}}{r}\left[2V^2+r(V^2-1)(\phi'+\lambda')+\frac{r \lambda'e^{-2\lambda}}{u^2+e^{-2\lambda}}\right] = 0,
\end{equation}
where $u = \dv*{r}{t}$, $V^2 = 4T/3(1+4T)$ with $T$ being the temperature of the fluid. The general idea of an accretion disc is that there is a Keplerian disc outside, where the matter has a minimal radial velocity. As the Keplerian disc terminates at a smaller distance from the black hole, the matter has a significant radial velocity, and it starts falling faster to the black hole. We assume that in this region, matter starts exhibiting spherical accretion flow. In other words, we assume that as the matter comes close enough to the black hole, it loses all its angular momentum, resulting in radial fall to the black hole (in reality, matter exhibits some angular momentum, even if sub-Keplerian). Figure~\ref{bondi_plot} shows the accretion and wind flows for the spherical accretion flow in $f(R)$ gravity, which are obtained by solving the Equation~\eqref{Eq: Bondi}. It is evident from the figure that the velocity is highly dependent on the outside temperature ($T_\mathrm{out}$), where matter starts behaving as a spherical accretion flow. The flow reaches the speed of light at the event horizon. As the event horizon shifts in the presence of $f(R)$ gravity, the flow reaches this maximum speed at different radii for different values of $\beta$. However, since the solution is asymptotically flat, both the accretion and wind curves match the well-known case for the Schwarzschild solution far from the black hole.

\section{Conclusions}
$f(R)$ gravity is one of the best bets to replace GR, and researchers have already proposed different models to explain various exciting phenomena related to cosmology and compact objects. For more than a decade, various researchers have put effort into obtaining the vacuum solution of $f(R)$ gravity. However, none of those solutions are asymptotically flat, which means they do not reduce to the Minkowski metric as one moves very far from any gravity source, and hence they may not be applicable to astrophysical phenomena. In this article, we have argued that asymptotically flat vacuum spacetime is also possible in $f(R)$ gravity. We have shown here that the vacuum solution in $f(R)$ gravity is not the Schwarzschild solution, which indicates the violation of Birkhoff's theorem in the presence of the modified gravity.   Since the solution is no longer the Schwarzschild solution, the quantities related to the event horizon and marginal orbits also alter, which further affects the spherical accretion flow around a black hole.

\bibliographystyle{MG_proceedings}
\bibliography{bibliography}

\begin{thebibliography}{10}

\bibitem{1980PhLB...91...99S}
A.~A. {Starobinsky}, {A new type of isotropic cosmological models without
  singularity}, {\em Physics Letters B} {\bf 91}, 99 (March 1980).

\bibitem{2014PhRvD..89j3509A}
A.~V. {Astashenok}, S.~{Capozziello} and S.~D. {Odintsov}, {Maximal neutron
  star mass and the resolution of the hyperon puzzle in modified gravity}, {\em
  \prd} {\bf 89}, p. 103509 (May 2014).

\bibitem{2018JCAP...09..007K}
S.~{Kalita} and B.~{Mukhopadhyay}, {Modified Einstein's gravity to probe the
  sub- and super-Chandrasekhar limiting mass white dwarfs: a new perspective to
  unify under- and over-luminous type Ia supernovae}, {\em \jcap} {\bf 9}, p.
  007 (September 2018).

\bibitem{2021ApJ...909...65K}
S.~{Kalita} and B.~{Mukhopadhyay}, {Gravitational Wave in f(R) Gravity:
  Possible Signature of Sub- and Super-Chandrasekhar Limiting-mass White
  Dwarfs}, {\em \apj} {\bf 909}, p.~65 (March 2021).

\bibitem{2006PhRvD..74f4022M}
T.~{Multam{\"a}ki} and I.~{Vilja}, {Spherically symmetric solutions of modified
  field equations in $f(R)$ theories of gravity}, {\em \prd} {\bf 74}, p.
  064022 (September 2006).

\bibitem{2007CQGra..24.2153C}
S.~{Capozziello}, A.~{Stabile} and A.~{Troisi}, {Spherically symmetric
  solutions in $f(R)$ gravity via the Noether symmetry approach}, {\em
  Classical and Quantum Gravity} {\bf 24}, 2153 (April 2007).

\bibitem{2010CQGra..27p5008C}
S.~{Capozziello}, M.~{De Laurentis} and A.~{Stabile}, {Axially symmetric
  solutions in $f(R)$-gravity}, {\em Classical and Quantum Gravity} {\bf 27},
  p. 165008 (August 2010).

\bibitem{2019EPJC...79..877K}
S.~{Kalita} and B.~{Mukhopadhyay}, {Asymptotically flat vacuum solution in
  modified theory of Einstein's gravity}, {\em European Physical Journal C}
  {\bf 79}, p. 877 (October 2019).

\bibitem{2010LRR....13....3D}
A.~{De Felice} and S.~{Tsujikawa}, {$f(R)$ Theories}, {\em Living Reviews in
  Relativity} {\bf 13}, p.~3 (June 2010).

\bibitem{1952MNRAS.112..195B}
H.~{Bondi}, {On spherically symmetrical accretion}, {\em \mnras} {\bf 112}, p.
  195  (1952).

\bibitem{1972Ap&SS..15..153M}
F.~C. {Michel}, {Accretion of Matter by Condensed Objects}, {\em \apss} {\bf
  15}, 153 (January 1972).

\end{thebibliography}

\end{document}